\documentclass{article}

\usepackage{arxiv}

\usepackage[utf8]{inputenc} 
\usepackage[T1]{fontenc}    
\usepackage{hyperref}       
\usepackage{url}            
\usepackage{booktabs}       
\usepackage{amsfonts}       
\usepackage{nicefrac}       
\usepackage{microtype}      
\usepackage{lipsum}
\usepackage{graphicx}
\graphicspath{ {./images/} }

\title{Smartphone-Based Food Traceability System Using NoSQL Database}

\author{
 Muhammad Syafrudin \\
  Department of Artificial Intelligence and Data Science\\
  Sejong University\\
  Seoul, Republic of Korea \\
  \texttt{udin@sejong.ac.kr} \\
   \And
 Ganjar Alfian \\
  Department of Electrical Engineering and Informatics\\
  Vocational College, Universitas Gadjah Mada\\
  Yogyakarta, Indonesia \\
  \texttt{ganjar.alfian@ugm.ac.id} \\
  \And
 Norma Latif Fitriyani \\
  Department of Artificial Intelligence and Data Science\\
  Sejong University\\
  Seoul, Republic of Korea \\
  \texttt{norma@sejong.ac.kr} \\
}

\begin{document}
\maketitle
\begin{abstract}
With growing consumer health awareness, ensuring food safety and quality throughout the supply chain is crucial, particularly for perishable goods. Contamination can occur during production, processing, or distribution, making real-time monitoring essential. This study proposes an affordable Smartphone-based food traceability system (FTS) that utilizes RFID technology and smartphone sensors. A smartphone-based RFID reader tracks products, while integrated sensors monitor temperature, humidity, and location during storage and transport. The system is assessed in the kimchi supply chain in Korea, providing real-time data to both managers and consumers. It offered comprehensive product tracking, including temperature and humidity records, ensuring transparency and safety. Compared to traditional methods, the proposed system demonstrated improved efficiency in handling large volumes of data while maintaining accurate traceability. The results highlight its potential for enhancing food safety and quality across supply chains. 
\end{abstract}


\section{Introduction}
Food safety has become a critical concern for all stakeholders involved in food production, driven by increasing consumer awareness regarding health and safety. As a result, the food industry faces growing pressure to enhance quality assurance, maintain product integrity, and ensure transparency throughout the entire food supply chain \cite{1}. Ensuring food safety and quality across the perishable food supply chain (PFSC) requires reliable systems to track and trace food products efficiently. A traceability system serves as a key solution by providing comprehensive, real-time information on food products, thereby reinforcing safety and quality standards \cite{2,3}. In this context, the Internet of Things (IoT) plays a vital role in facilitating communication between devices, with technologies such as RFID (Radio Frequency Identification) being widely adopted for tracking and monitoring products in perishable food supply chains (PFSCs) \cite{4,5,6}.

Building upon the increasing reliance on IoT technologies, IoT sensors, including temperature and humidity monitors, are increasingly used to enhance food safety by enabling real-time environmental condition monitoring \cite{7}. Given that temperature is a critical factor in determining the safety and quality of perishable foods, continuous monitoring of temperature during transportation and storage is essential \cite{8,9}. Smartphones, serving as gateways for IoT data, are capable of monitoring environmental conditions and providing real-time location data via embedded GPS sensors, further improving supply chain management and food safety practices \cite{10,11,12}. Smartphones equipped with a range of sensors serve as powerful tools for monitoring agricultural environments and improving crop management \cite{13}. As central hubs for data collection, processing, and real-time decision-making in precision agriculture, smartphones enhance the ability of farmers to optimize their activities and ensure food safety across the supply chain. Research demonstrates that smartphone-based environmental monitoring systems can effectively track conditions under which perishable goods are transported and stored, allowing for optimal preservation of food quality and providing real-time data to supply chain stakeholders \cite{14}. Furthermore, smartphones are increasingly recognized as cost-effective tools for traceability systems, enabling producers and consumers to track and verify the origin and transportation of food products, thus fostering transparency and safety in perishable food supply chains \cite{15}.

Expanding on these innovations, this study proposes a smartphone-based food traceability system (FTS) that combines RFID technology with smartphone-based sensors to enhance food safety and quality in PFSCs. By employing RFID for product tracking and traceability, the system reduces the risk of counterfeiting and prevents the distribution of substandard food products. Additionally, temperature-humidity sensors and GPS integrated into smartphones monitor environmental conditions throughout the supply chain. This technological integration ensures the quality and safety of agricultural products, allowing customers to verify the product's integrity at any stage, thereby meeting the industry's growing demand for transparency and consumer trust.

The study contributions are summarized as follows:
\begin{itemize}
\item Development of a Smartphone-based FTS for PFSC.
\item Integration of RFID technology for enhanced tracking and tracing capabilities.
\item Implementation of smartphone-based sensors for real-time monitoring of location data, humidity, and temperature.
\item Improved food safety and transparency across the entire supply chain.
\item Enhanced consumer trust through verifiable food quality assurance. 
\end{itemize}

\section{Materials and Methods}
\subsection{Smartphone-Based FTS Design}
The proposed Smartphone-based FTS is designed to guarantee the safety and quality of food products in the PFSC. The system employs a combination of RFID handheld readers and smartphone-based sensors to track and monitor food products across the supply chain.

\begin{figure}
  \centering
  \includegraphics{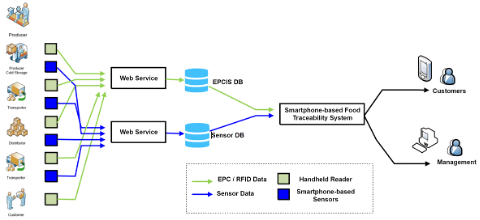}
  \caption{Architecture of smartphone-based food traceability system (FTS).}
  \label{fig:fig1}
\end{figure}

In this study, we focus on the kimchi supply chain, which includes the producer, transporter, distributor, and consumer (school canteens), as can illustrated in Fig. \ref{fig:fig1}. The process begins with the producer, who stores the kimchi in cold storage after production. Once the product is ready for shipment, the transporter moves it to the distribution center and eventually delivers it to the final consumers. At each stage of the supply chain, partners use RFID handheld readers to capture the electronic product code (EPC) of the kimchi, which is stored in the EPC Information Service (EPCIS) \cite{16,17} database. This EPCIS database follows the GS1 standard, allowing stakeholders to exchange detailed information regarding the physical movement and status of products. The system provides answers to key questions about the product's journey, such as "what," "where," "when," and "why," meeting the needs of both regulatory and consumer requirements by providing accurate and comprehensive product information.

In addition to RFID tracking, the system leverages smartphones equipped with add-on temperature-humidity sensors and embedded GPS modules. These sensors monitor environmental conditions such as temperature and humidity and track the location of the product throughout the supply chain. The data collected by the sensors is transmitted from the smartphone to the server and stored in the sensor database. Finally, the Smartphone-based traceability system presents users with a complete history of the product, including humidity, temperature, and location data (based on GPS coordinates), ensuring full transparency and enhancing food safety throughout the supply chain.

\subsection{The Proposed Smartphone-Based FTS Implementation}
\begin{figure}
  \centering
  \includegraphics[width=0.5\textwidth]{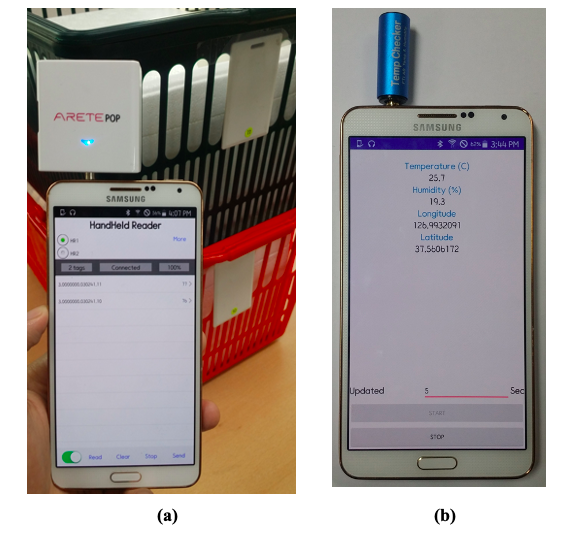}
  \caption{Implementation of smartphone-based FTS: (a) handheld reader and (b) sensors.}
  \label{fig:fig2}
\end{figure}

At the kimchi production facility, freshly made kimchi were packed into specialized kimchi boxes on the production line, with passive RFID tags attached to each box. These passive card-type tags were made of PVC and measured 86×54×1.8 mm. Operating at a frequency of 902–928 MHz and following the EPC Gen2 (ISO 18000-6c) protocol \cite{18}, these tags enabled identification and tracking. The RFID data was captured using an Arete Pop Dongle RFID Reader \cite{19} paired with a custom Android application. This compact reader, designed for short-range use, is compatible with smartphones via the 3.5 mm headphone jack and utilizes the PHYCHIPS PR9200 RFID chip. The reader transfers RFID data to the connected smartphone, which then forwards it to a web service over the internet. Staff used this handheld reader during storage and transportation to monitor and track the kimchi boxes. Fig. \ref{fig:fig2}(a) and \ref{fig:fig2}(b) illustrates an example of the handheld reader and sensor in use, respectively.

\begin{figure}
  \centering
  \includegraphics[width=.8\textwidth]{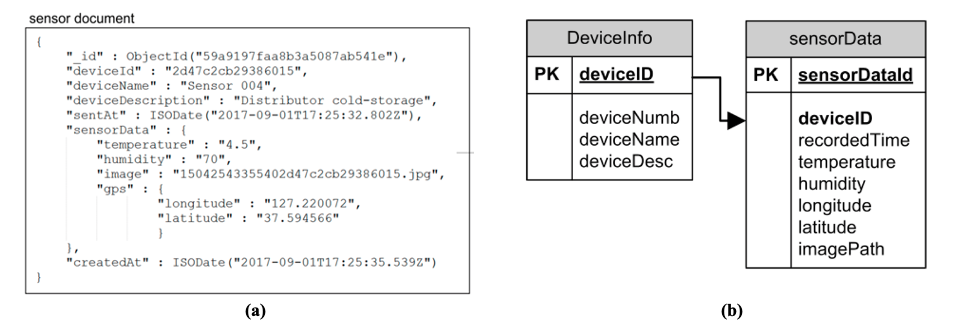}
  \caption{An illustration of (a) a Smartphone-based generated sensor record in a NoSQL (MongoDB version 3.6) and (b) a sensor schema in a SQL (MySQL version 5.7) database \cite{10}.}
  \label{fig:fig3}
\end{figure}

The sensor data generated by Smartphone gateways is typically large, unstructured, and continuously produced. This project examines two data schemas, one utilizing NoSQL (MongoDB version 3.6) and the other employing SQL  (MySQL version 5.7) databases. In the case of the NoSQL database, Kang et al. (2016) implemented an embedding scheme for the sensor data repository, given that the data remains largely unmodified (raw data) \cite{20}. The current study confirms that this embedding approach is well-suited for managing extensive sensor data repositories, as it ensures efficient read and write operations. Consequently, we adopt an embedding scheme-based sensor data repository, as illustrated in Fig. \ref{fig:fig3}(a). Each sensor document in this repository includes a unique smartphone device ID, a timestamp, and sensor data organized within a sub-document.

For the SQL-based sensor data repository, we establish two tables: DeviceInfo and sensorData, as depicted in Fig. \ref{fig:fig3}(b). The DeviceInfo table captures information about the sensor devices, with the deviceID column designated as the primary key. Meanwhile, the sensorData table serves as the relational component, recording details such as temperature, humidity, GPS coordinates (longitude and latitude), and images from each sensor device, organized by timestamp. In this structure, the deviceID column in the sensorData table acts as a foreign key, linking back to the deviceID in the DeviceInfo table.

To monitor environmental conditions during the supply chain, temperature-humidity sensors and GPS data were integrated through smartphones. For this, the Smart Temp Checker FTC-001 device \cite{21} was employed, which was designed for individual smartphone use. This device also connects via the 3.5 mm headphone jack to capture temperature and humidity data and forward it to the smartphone. Similar to the RFID reader, the Smart Temp Checker provides an Android API, allowing the app to receive sensor data and transmit it to a remote server via the Internet. Fig. \ref{fig:fig2}(b) depicts the smartphone-based sensor in operation. These smartphone-based sensors were installed at various points along the supply chain—such as at the production facility, with transporters, and within cold storage facilities at distributors—to collect comprehensive temperature, humidity, and GPS data for each location throughout the distribution process.

\section{Result and Discussion}
\subsection{Performance Analysis of Smartphone-Based FTS}

\begin{figure}
  \centering
  \includegraphics[width=.9\textwidth]{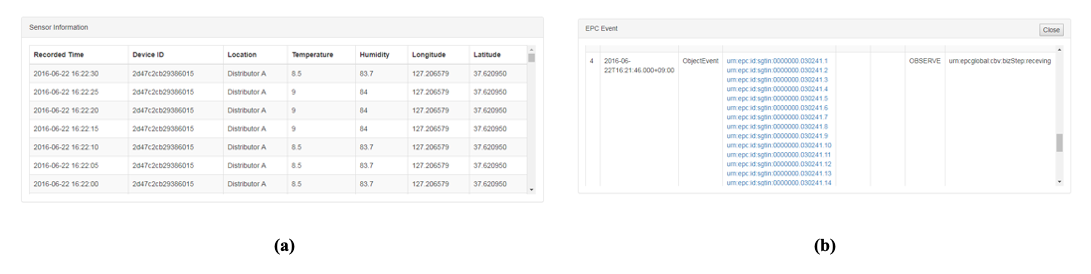}
  \caption{The webpage interface of smartphone-based FTS: (a) gathered sensor data and (b) product history.}
  \label{fig:fig4}
\end{figure}

This study developed a Smartphone-based FTS, offering a user-friendly method for tracking and tracing the history of food products. The server-side application, built on the Express framework for Node.js, utilizes a NoSQL MongoDB version 3.6 database to store sensor data.
Monitoring temperature and humidity is crucial for verifying the quality of perishable agricultural products. Fig. \ref{fig:fig4}(a) displays sensor data, including temperature, humidity, longitude, and latitude, with time stamps and device IDs. In this example, sensor data was collected every 5 seconds from a smartphone-based sensor within the kimchi supply chain network.
The system displays product history, including object ID (EPC), time, location, and business context. Fig. \ref{fig:fig4}(b) illustrates a screenshot showing the food product's historical data.

System performance was evaluated using two metrics: server response time and device response time. Server response time refers to the duration needed to store sensor data in the database, while device response time is the time required for a device to read the products. Server response time was assessed through simulations where a Java program, running on a client computer with a CPU of 2.53 GHz and 16 GB RAM, generated and stored varying amounts of sensor data in the database. The IBM System x3250 M3 was used as the server. For device response time, the performance of RFID handheld readers was compared to QR code readers. The average response times were recorded from multiple readings.

\begin{figure}
  \centering
  \includegraphics[width=.8\textwidth]{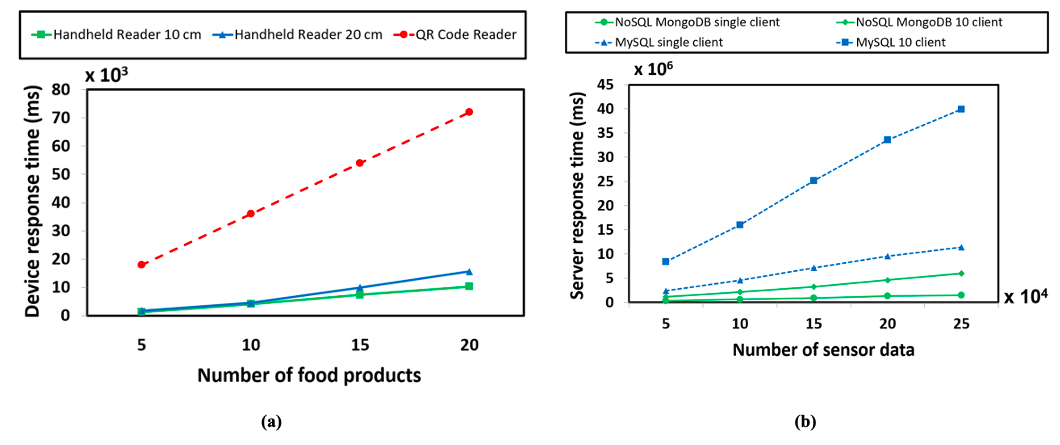}
  \caption{Performance analysis of smartphone-based FTS: (a) device and (b) server response time under different scenarios.}
  \label{fig:fig5}
\end{figure}

Fig. \ref{fig:fig5}(a) compares the performance of RFID handheld readers with QR code readers. Two RFID reading distances, 10 cm, and 20 cm, were evaluated. Results indicated that the RFID handheld reader had a shorter response time compared to the QR code reader and performed more efficiently, especially at shorter reading distances. As the number of objects increased, the RFID reader maintained a faster reading rate than the QR code reader.

Fig. \ref{fig:fig5}(b) shows that as the volume of sensor data increased, the server's response time also rose. Additionally, a higher number of clients led to longer server response times due to increased data storage demands. The NoSQL MongoDB-based (version 3.6) system outperformed MySQL (version 5.7) in this context. MongoDB outperforms MySQL in several key areas, primarily due to its inherent design and architecture. One of the most significant advantages of MongoDB is its schema-less nature, allowing for flexible data models that can accommodate unstructured or semi-structured data. This flexibility enables developers to quickly adapt the database structure to changing business requirements without the need for extensive migrations or downtime. For food traceability systems, which often involve diverse data types (such as sensor readings, transaction logs, and product details), this adaptability is essential for rapid implementation and iterative improvements.

Additionally, MongoDB's ability to handle high-throughput read and write operations makes it particularly suited for applications with dynamic workloads. The database is optimized for horizontal scaling, meaning it can easily distribute data across multiple nodes. This characteristic becomes especially beneficial during peak operational periods, such as harvest seasons or product launches, where sudden spikes in data input and retrieval requests can occur. As a result, MongoDB can maintain consistent performance under high loads, ensuring that the system remains responsive and capable of providing real-time insights into food safety and quality.
On the other hand, while MySQL excels in maintaining data integrity through its support for ACID (Atomicity, Consistency, Isolation, Durability) properties, it can experience performance bottlenecks when handling large volumes of data or complex queries. As data grows, the structured schema of MySQL requires careful optimization, such as indexing, to ensure efficient query execution. This reliance on a fixed schema and complex join operations can lead to slower performance in scenarios involving extensive data relationships, such as tracking the provenance of food products through various stages of the supply chain.
Furthermore, MySQL’s transaction handling can introduce overhead, particularly in write-intensive environments. Each transaction must adhere to strict consistency rules, which can slow down the insertion of new records, such as sensor data. In contrast, MongoDB’s eventual consistency model allows for faster writes, as it does not enforce immediate consistency across all nodes. This trade-off can be advantageous in applications where speed is prioritized over strict data integrity, allowing for more agile responses to real-time data inputs.

In summary, MongoDB offers several performance advantages over MySQL in the context of food traceability systems, including its flexible schema design, high scalability, and ability to handle large volumes of data with minimal latency. However, the choice between these databases should also consider the specific operational requirements, such as the need for data integrity and the complexity of queries, which may favor the structured approach of MySQL. Ultimately, a hybrid database strategy that leverages the strengths of both NoSQL and relational databases could provide an optimal solution for managing diverse data in food traceability applications.

\subsection{Trade-offs Between NoSQL and Relational Databases Analysis of Smartphone-Based FTS}
Expanding on the performance analysis of the proposed smartphone-based food traceability system, it is essential to consider the trade-offs between using NoSQL databases, such as MongoDB, and traditional relational databases, such as MySQL, particularly under varying loads.
NoSQL databases like MongoDB are designed to handle large volumes of unstructured or semi-structured data, making them suitable for applications with dynamic data models, such as those found in food traceability systems. MongoDB’s flexibility allows for rapid data ingestion and scaling, which can be particularly beneficial when dealing with fluctuating data loads that occur during peak seasons or unexpected surges in demand. Its schema-less architecture enables the system to accommodate new types of data without requiring significant reconfiguration, thus facilitating quick adaptations to evolving business needs.

However, this flexibility comes with trade-offs. While MongoDB excels at read and write operations with high throughput, it may sacrifice some consistency and transactional integrity compared to traditional relational databases. MySQL, with its structured schema and support for ACID (Atomicity, Consistency, Isolation, Durability) properties, ensures robust data integrity and complex querying capabilities. This can be particularly advantageous in scenarios requiring rigorous compliance with food safety regulations and standards. Under high loads, MySQL may experience performance bottlenecks due to its reliance on structured data and the overhead associated with maintaining transactional integrity.

Conversely, in scenarios where rapid data retrieval and scalability are prioritized, NoSQL databases may outperform relational databases. For instance, during peak operational periods, such as harvest seasons or promotional events, the ability of MongoDB to horizontally scale by distributing data across multiple nodes can ensure consistent performance without significant degradation. However, this scalability can lead to challenges in data consistency if not managed properly, potentially impacting real-time decision-making capabilities within the traceability system.

In summary, the choice between NoSQL databases like MongoDB and traditional relational databases like MySQL should be guided by the specific requirements of the food traceability system, including data structure, consistency needs, and expected load patterns. A hybrid approach, leveraging the strengths of both database types, may also be considered to optimize performance, ensuring the system remains agile and responsive under varying conditions while maintaining data integrity and compliance with industry standards.

\subsection{Cost Analysis of Smartphone-Based FTS}
The cost of implementing a Smartphone-based FTS involves several key components:
\begin{itemize}
\item RFID Dongle Reader: The Arete Pop RFID dongle reader costs approximately \$200. This reader has a short-range capability (up to 1 meter, depending on the RFID tags) and includes a rechargeable battery with a 2-hour continuous use capacity.
\item Smartphones: Smartphones used in the system range from \$100 to \$300. Lower-cost models are sufficient, provided they can connect to the internet to communicate with the server.
\item System Software and Hardware: Utilizing open-source software (OSS) can lead to significant cost savings in software development, product quality, and license fees \cite{22,23,24}. Our system is built on OSS, which is economically advantageous for implementation and integration. For hardware, we employed the IBM System x3250M3 server, which costs between \$1,000 and \$2,000, offering a cost-effective and compact solution.
\item Smartphone-Based Sensors: The Smart Temp Checker FTC-001 sensor, used for monitoring temperature and humidity, costs between \$20 and \$30 per unit. This sensor features a built-in battery that supports continuous operation for up to 250 hours, with temperature and humidity measurement errors of ±0.5°C and ±2.5\%, respectively.
\item RFID Passive Tags: Card-type passive RFID tags are priced at less than \$1 each. In this study, these tags are affixed to kimchi shipping boxes. The boxes are returned to the producer after delivery, making them reusable and reducing overall costs.
\end{itemize}

In summary, the costs associated with implementing the Smartphone-based traceability system are manageable and support scalability and sustainability. By using a combination of affordable components and leveraging OSS, the system remains both cost-effective and efficient.

\subsection{Study Limitations}
While the proposed smartphone-based food traceability system (FTS) offers significant advantages in enhancing food safety and quality in perishable food supply chains, it is essential to consider potential limitations.
One primary concern is the scalability of the solution, particularly in larger supply chains. As the size and complexity of a supply chain increase, the volume of data generated by numerous sensors and RFID tags can become overwhelming. Efficiently managing and processing this vast amount of data requires robust infrastructure and advanced data analytics capabilities. Without adequate scalability, the system may struggle to maintain performance and reliability, potentially hindering its effectiveness in ensuring food safety.

Additionally, deploying the FTS across different regions poses unique challenges. Variations in local regulations, technological infrastructure, and consumer acceptance can affect the implementation and operation of the system. In regions with limited access to smartphone technology or unreliable internet connectivity, the effectiveness of the system may be compromised. Furthermore, cultural differences in food handling and safety practices can impact the adoption and use of the traceability features offered by the FTS.

Another limitation pertains to the integration of the system with existing supply chain processes. For successful adoption, stakeholders throughout the supply chain must be willing to cooperate and share data. Resistance from participants due to concerns about privacy, competition, or the perceived complexity of the system could impede its implementation.
Lastly, the reliance on technology for monitoring and data collection introduces potential vulnerabilities. Technical issues, such as sensor malfunctions or software bugs, could lead to inaccuracies in data reporting, undermining trust in the traceability system. Addressing these limitations through thorough planning, stakeholder engagement, and ongoing support will be crucial for the successful deployment and long-term viability of the smartphone-based FTS.

\section{Conclusion and Future Work}
The proposed Smartphone-based FTS significantly enhances food quality and safety by enabling precise tracking, tracing, and monitoring of perishable products' temperature and humidity. By integrating RFID technology and smartphone-based sensors, the system demonstrates substantial scalability. As the volume of sensor data and the number of objects increases, the system processes data more efficiently compared to traditional methods. This capability allows for better tracking and tracing of products, reducing the risk of counterfeiting and ensuring high-quality food throughout the supply chain.

In conclusion, the proposed smartphone-based food traceability system offers significant potential for enhancing food safety and quality. Future work should focus on integrating data mining techniques to improve decision-making within the system. Specifically, implementing predictive analytics can help anticipate risks in the supply chain, while anomaly detection algorithms can identify unusual patterns in real-time to enable swift corrective actions. Additionally, assessing supplier performance through historical data analysis will facilitate data-driven procurement decisions, and analyzing consumer behavior can align offerings with market demand. Furthermore, exploring the integration of data mining with blockchain technology can enhance transparency and traceability, providing immutable records that foster trust among stakeholders. These advancements will not only streamline operations but also strengthen food safety and consumer confidence in the supply chain.

\bibliographystyle{unsrt}  


\begin{thebibliography}{1}

\bibitem{1}
J. Trienekens and P. Zuurbier
\newblock Quality and safety standards in the food industry, developments and challenges
\newblock International Journal of Production Economics, vol. 113, no. 1, pp. 107–122, May 2008, doi: 10.1016/j.ijpe.2007.02.050.

\bibitem{2}
A. Regattieri, M. Gamberi, and R. Manzini
\newblock Traceability of food products: General framework and experimental evidence
\newblock Journal of Food Engineering, vol. 81, no. 2, pp. 347–356, Jul. 2007, doi: 10.1016/j.jfoodeng.2006.10.032.

\bibitem{3}
M. M. Aung and Y. S. Chang
\newblock Traceability in a food supply chain: Safety and quality perspectives
\newblock Food Control, vol. 39, pp. 172–184, May 2014, doi: 10.1016/j.foodcont.2013.11.007.

\bibitem{4}
L. Wang, S. K. Kwok, and W. H. Ip
\newblock A radio frequency identification and sensor-based system for the transportation of food
\newblock Journal of Food Engineering, vol. 101, no. 1, pp. 120–129, Nov. 2010, doi: 10.1016/j.jfoodeng.2010.06.020.

\bibitem{5}
G. Alfian et al. 
\newblock Improving efficiency of RFID-based traceability system for perishable food by utilizing IoT sensors and machine learning model
\newblock Food Control, vol. 110, p. 107016, Apr. 2020, doi: 10.1016/j.foodcont.2019.107016.

\bibitem{6}
E. Abad et al.
\newblock RFID smart tag for traceability and cold chain monitoring of foods: Demonstration in an intercontinental fresh fish logistic chain
\newblock Journal of Food Engineering, vol. 93, no. 4, pp. 394–399, Aug. 2009, doi: 10.1016/j.jfoodeng.2009.02.004.

\bibitem{7}
Y. Bouzembrak, M. Klüche, A. Gavai, and H. J. P. Marvin
\newblock Internet of Things in food safety: Literature review and a bibliometric analysis
\newblock Trends in Food Science \& Technology, vol. 94, pp. 54–64, Dec. 2019, doi: 10.1016/j.tifs.2019.11.002.

\bibitem{8}
M. M. Aung and Y. S. Chang 
\newblock Temperature management for the quality assurance of a perishable food supply chain
\newblock Food Control, vol. 40, pp. 198–207, Jun. 2014, doi: 10.1016/j.foodcont.2013.11.016.

\bibitem{9}
M. Thakur and E. Forås
\newblock EPCIS based online temperature monitoring and traceability in a cold meat chain
\newblock Computers and Electronics in Agriculture, vol. 117, pp. 22–30, Sep. 2015, doi: 10.1016/j.compag.2015.07.006.

\bibitem{10}
G. Alfian, M. Syafrudin, and J. Rhee 
\newblock Real-Time Monitoring System Using Smartphone-Based Sensors and NoSQL Database for Perishable Supply Chain
\newblock Sustainability, vol. 9, no. 11, p. 2073, Nov. 2017, doi: 10.3390/su9112073.

\bibitem{11}
G. Aloi et al.
\newblock Enabling IoT interoperability through opportunistic smartphone-based mobile gateways
\newblock Journal of Network and Computer Applications, vol. 81, pp. 74–84, Mar. 2017, doi: 10.1016/j.jnca.2016.10.013.

\bibitem{12}
S. Sharma, J. Raval, and B. Jagyasi
\newblock Mobile sensing for agriculture activities detection
\newblock in 2013 IEEE Global Humanitarian Technology Conference (GHTC), San Jose, CA, USA: IEEE, Oct. 2013, pp. 337–342. doi: 10.1109/GHTC.2013.6713707.

\bibitem{13}
P. Rajak, A. Ganguly, S. Adhikary, and S. Bhattacharya 
\newblock Internet of Things and smart sensors in agriculture: Scopes and challenges
\newblock Journal of Agriculture and Food Research, vol. 14, p. 100776, Dec. 2023, doi: 10.1016/j.jafr.2023.100776.

\bibitem{14}
V. Sathiya, K. Nagalakshmi, K. Raju, and R. Lavanya
\newblock Tracking perishable foods in the supply chain using chain of things technology
\newblock Sci Rep, vol. 14, no. 1, p. 21621, Sep. 2024, doi: 10.1038/s41598-024-72617-3.

\bibitem{15}
K. Lin, D. Chavalarias, M. Panahi, T. Yeh, K. Takimoto, and M. Mizoguchi 
\newblock Mobile-based traceability system for sustainable food supply networks
\newblock Nat Food, vol. 1, no. 11, pp. 673–679, Nov. 2020, doi: 10.1038/s43016-020-00163-y.

\bibitem{16}
J. Ahn, H. Gaza, J. Lee, H. Kim, and J. Byun
\newblock Oliot EPCIS: An open-source EPCIS 2.0 system for supply chain transparency
\newblock SoftwareX, vol. 23, p. 101477, Jul. 2023, doi: 10.1016/j.softx.2023.101477.

\bibitem{17}
H. L. Gaza and J. Byun 
\newblock tracES++: Applying Incremental Computation to Temporal Information Diffusion Analysis for Online Object Traceability
\newblock IEEE Access, vol. 12, pp. 78811–78824, 2024, doi: 10.1109/ACCESS.2024.3394543.

\bibitem{18}
“EPC C-1 G-2 / ISO 18000-6C RFID IC,”  
\newblock EPC C-1 G-2 / ISO 18000-6C RFID IC. Accessed: Aug. 13, 2024. [Online]
\newblock Available: https://www.emmicroelectronic.com/sites/default/files/products/datasheets/4124-ds.pdf

\bibitem{19}
“ARETE POP | tradekorea,” 
\newblock ARETE POP | tradekorea. Accessed: Aug. 26, 2024. [Online]
\newblock Available: https://web.archive.org/web/20240902085556/https://tradekorea.com/product/detail/P341461/ARETE-POP.html

\bibitem{20}
Y.-S. Kang, I.-H. Park, J. Rhee, and Y.-H. Lee
\newblock MongoDB-Based Repository Design for IoT-Generated RFID/Sensor Big Data
\newblock IEEE Sensors J., vol. 16, no. 2, pp. 485–497, Jan. 2016, doi: 10.1109/JSEN.2015.2483499.


\bibitem{21}
“Denshine® FTLAB FTC-001 SmartLab Smart Temperature Humidity Checker (Blue),” 
\newblock Denshine® FTLAB FTC-001 SmartLab Smart Temperature Humidity Checker (Blue). 
\newblock Accessed: Aug. 26, 2024. [Online]. Available: https://web.archive.org/web/20240902090016/https://www.amazon.ca/Denshine\%C2\%AE-FTC-001-SmartLab-Temperature-Humidity/dp/B01D2XI4M2

\bibitem{22}
M. Syafrudin, N. Fitriyani, D. Li, G. Alfian, J. Rhee, and Y.-S. Kang 
\newblock An Open Source-Based Real-Time Data Processing Architecture Framework for Manufacturing Sustainability
\newblock Sustainability, vol. 9, no. 11, p. 2139, Nov. 2017, doi: 10.3390/su9112139.

\bibitem{23}
S. A. Ajila and D. Wu
\newblock Empirical study of the effects of open source adoption on software development economics
\newblock Journal of Systems and Software, vol. 80, no. 9, pp. 1517–1529, Sep. 2007, doi: 10.1016/j.jss.2007.01.011.

\bibitem{24}
K. Ven and J. Verelst
\newblock The Organizational Adoption of Open Source Server Software by Belgian Organizations
\newblock in Open Source Systems, vol. 203, E. Damiani, B. Fitzgerald, W. Scacchi, M. Scotto, and G. Succi, Eds., in IFIP International Federation for Information Processing, vol. 203. , Boston, MA: Springer US, 2006, pp. 111–122. doi: 10.1007/0-387-34226-5\_11.

\end{thebibliography}

\end{document}